\documentstyle[emulateapj]{article}

\begin{document}
\title{Galaxy Luminosity Function in Groups}

\author{Hern\'an Muriel, Carlos A. Valotto and Diego G. Lambas}

\affil{Grupo de Investigaciones en Astronom\'{\i}a Te\'orica y
Experimental (IATE)}

\affil{Observatorio Astron\'omico de C\'ordoba, Laprida 854, 5000
C\'ordoba, Argentina \\}
\affil{CONICET, Buenos Aires, Argentina\\
       hernan@oac.uncor.edu, val@oac.uncor.edu, dgl@oac.uncor.edu}

\begin{abstract}

We compute the luminosity function of galaxies in groups. 
The data consists in two samples of groups of galaxies
selected in distance and redshift space 
comprising a total number of 66 groups.
The assignment of galaxies to the groups were derived
from the Edinburgh-Durham Southern Galaxy Catalog, 
considering a limiting apparent magnitude $m_{lim}=19.4$.

We find a flat faint end of the mean galaxy luminosity function in groups 
in contrast to the mean galaxy LF in clusters where
a large relative number of faint galaxies is present. 
We find that 
a Schechter function with parameters $M^{*}\simeq -19.6\pm 0.2,$ $\alpha
\simeq -1.0\pm $ $0.2$ provides a satisfactory fit to the galaxy
LF of our total sample of groups   
in the range of absolute magnitudes $-22<M<-16$.
\end{abstract}

\section{INTRODUCTION}

Several works have attempted the identification of groups of galaxies
in 2 dimensions using catalogs with angular positions (see for instance
\cite{tur76}).
A major problem with such identifications is the low density contrast 
of these systems and therefore redshift surveys are essential to provide
reliable physical groups. 
Nevertheless, most of redshift surveys are limited to bright
apparent magnitudes allowing  only the identification of the brightest
members. This fact make difficult the  determinations
 of the faint end of the luminosity function (hereafter LF)
 of galaxies in groups. 
The galaxy LF in clusters and nearby groups has
been studied by several authors (\cite{oem74}, \cite{dre78}, \cite{oeg86}, 
\cite{oeg87}, \cite{gud91}, \cite{wil90}, \cite{gar91}, \cite{fer91},
\cite{gar92}, \cite{lop96}, \cite{gai97}, \cite{val97}, \cite{tren97}). Of
particular importance is the possible universal character of the galaxy LF
early suggested by \cite{abe62} and \cite{abe75} since it may serve to pose 
constraints to models of galaxy formation as well as testing the importance of
environment in the related astrophysical processes.
Interactions play a fundamental role in 
galaxy evolution (see for instance \cite{pos84}) 
being particularly relevant in small galaxy associations.
Moreover, in contrast to clusters, groups have a low galaxy
velocity dispersion, gas density and temperature
which may cause a significantly different evolution of galaxies 
in groups and clusters.
Groups of galaxies are therefore suitable systems to provide a useful 
observational insight on different physical phenomena related to environment.

The field galaxy LF has been determined in several works, see for instance
\cite{lov92}, and more recently
\cite{mar94} and \cite{lin96}. These authors provide Schechter function fits
(\cite{sch76}) and in spite of discrepancies in the value of the
parameter $M^{*}$, both determinations are consistent with a flat LF at the
faint end, $\alpha \simeq -1.$ On the other hand, several studies of the
galaxy luminosity function in clusters show a much steeper galaxy LF at
the faint end, $\alpha \leq -1.4$ (e.g. \cite{fer91}, \cite{val97}, 
\cite{tren97}).

The universality of the galaxy LF in clusters has been seriously
questioned by \cite{lc95} and  \cite{lc97}.
From their sample of 45 Abell clusters with z$<$0.14
39 clusters show an increase of the fraction of dwarf galaxies consistent
with a double Schechter fit with $\alpha \simeq 1$ and $-2<\alpha <-1.4$ and only
7 can be suitable fitted with a single Schechter fit with $\alpha \simeq 1$.
This set of clusters with flat LF are characterized by an evolved
morphology of the cD type and are on
average massive and gas rich. Nevertheless, \cite{driver}
and \cite{tren97} find rising LF in rich clusters
which suggest the complexity of this phenomenon.
In spite of the reported differences of the galaxy LF in clusters
the results show agreement with the
mean galaxy LF in clusters derived by
\cite{val97} which rises at
faint magnitudes consistent with a single Schechter fit with $\alpha \simeq -1.4$.
The tendency of poorer clusters to show a flatter galaxy LF ($\alpha \simeq -1.2$)
than richer clusters ($\alpha \simeq -1.5$)  was obtained by \cite{val97} 
dividing their sample
in two equal number of clusters. These results and \cite{lc97} findings
would not be necessarily inconsistent given that the 
 flat LF clusters ($\alpha \simeq -1.0$)
observed correspond to a small fraction ($\simeq 20\% $) 
of dynamically relaxed
systems where evolutionary process have an important role.
Other works provide useful studies of the
galaxy LF in different environments. For instance, \cite{gai97} develop an
important work on the luminosity function of galaxies in clusters.
Nevertheless, the background subtraction in this work is taken 
at 0.5 Mpc of the cluster center
which may contribute significantly to the resulting flat LF of the 
analysis given the expected large contamination of the
background by cluster members. 

Given the observed differences between cluster and 
field mean galaxy luminosity functions 
we analyze in this work the galaxy LF in moderate density associations.
In order to provide a suitable study of the effects of environment  
we use the same statistical procedures as in \cite{val97}
avoiding possible systematic effects due to the use of
different techniques.
In section 2 we describe the group and galaxy data used. 
Section 3 gives a description of the methods of analysis
adopted, the main results obtained and
error estimates through a Monte-Carlo method. In Section 4 we present
the main conclusions.

\section{DATA}

Willick et al. (1997) have assembled a homogeneous catalog of peculiar
velocity data (the Mark III catalog). This catalog includes a sample
of groups selected using redshifts and distances of the Mark III galaxies
with a grouping algorithm as described in \cite{wil96}. Although the
authors claim that the dynamical characteristics of the groups are 
in any sense well defined, this sample is suitable for our analysis 
since our aim is to study the galaxy LF in moderate galaxy overdensities.
In addition, completeness in group numbership is not required given that COSMOS
galaxies are used to provide a statistical
assignment to groups. The Mark III
group sample comprise a total number of 277 objects.    
This sample of groups of galaxies has redshift independent distance
estimates which enables us to study the galaxy LF for nearby groups 
of galaxies for which peculiar motions cannot be neglected.

Other samples of groups of galaxies without redshift independent
distance estimates was taken from different group catalogs 
(\cite{mai89} (MCL), \cite{fou92} (FGCP) and  \cite{gar93}). 
Garc\'{\i}a (1993) identify groups in the magnitude-limited 
(B$_{0}$ brighter than
14.0) sample of galaxies extracted from the Lyon-Meudon Extragalactic Database
(LEDA).
This author identify groups applying two different technique: the
hierarchical clustering method (\cite{mat78}) and the percolation method
(\cite{huc82}). Garc\'{\i}a identify 485 groups of at least three
members and detected by both technique.
MCL select groups of galaxies using the same technique applied by HG
in the CfA catalog. This technique is apply to the Southern Sky
Redshift Survey (\cite{dac88}). MCL identify 87 groups with more
than two members and mean velocities smaller than 8000 km s$^{-1}$.
FGCP list 246 groups of at least three members obtained by applying the
hierarchical algorithm to the Catalog of Principal Galaxies (\cite{pat89a},
\cite{pat89b}).
We have also included compact groups taken from \cite{hic82}).

We have restricted this compilation of groups without redshift independent
distance estimates to radial velocities 
$cz > 2500$ km s$^{-1}$. With this restriction the group peculiar 
velocities would not seriously
affect the estimates of absolute magnitudes due to peculiar motions.
Since we aim to study moderate galaxy density enhancements 
we have not considered groups within 0.5 Mpc $h^{-1}$  in projection to Abell
clusters of galaxies.
We have adopted a limiting absolute magnitudes 
$M_{lim}=-16$ since the signal-to-noise ratio for fainter absolute
magnitudes results too weak in our studies which impose a further
restriction to group distances.

Our final sample comprises 66 groups, 35 with distances derived from the
MarkIII catalog and 31 from our compilation of 
groups with redshift determinations from the literature
from which 3 are Hickson compact groups.
In figure 1 are displayed the distribution of distances for
the two subsamples of groups. 

\placefigure{fig-1}
\input epsf
\def\figureps[#1,#2]#3.{\bgroup\vbox{\epsfxsize=#2
    \hbox to \hsize{\hfil\epsfbox{#1}\hfil}}\vskip12pt
    \small\noindent Figure#3. \def\par{\endgraf\egroup\vskip12pt}}
\figureps[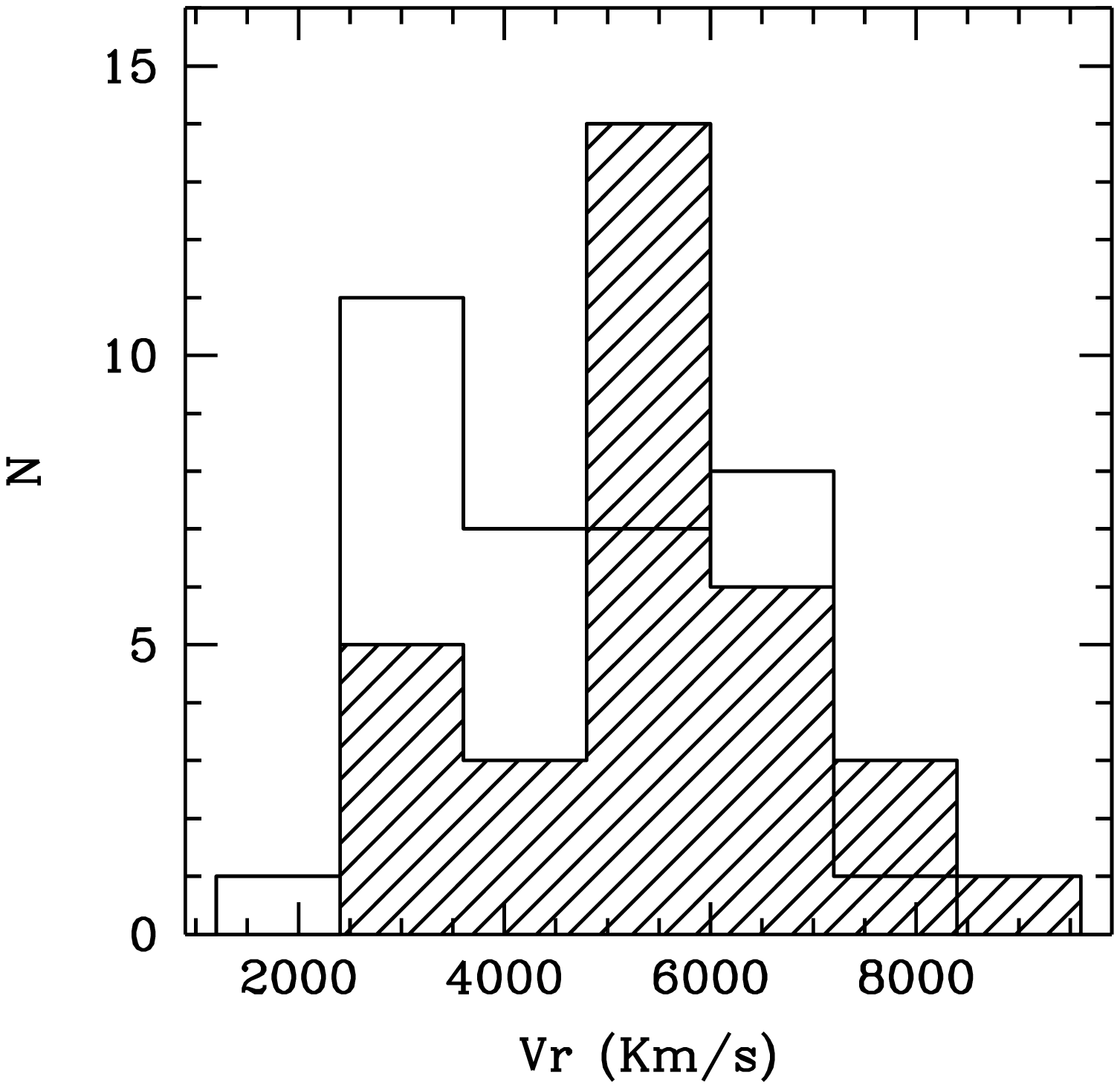,1.00\hsize] 1. Distribution of distances in km s$^{-1}$ 
corresponding to 35 MarkIII groups and 31 groups with radial
velocities (filled histogram) considered in our study.

The Edinburgh-Durham Southern Galaxy Catalog, hereafter COSMOS Survey
(\cite{hey89}), was used for the statistical assignment of
galaxies to the groups. This survey provides angular positions and
photographic magnitudes in the $b_{j}$ band for over two million galaxies. We
have restricted to the region $\delta <-10 \arcdeg$ given the 
lower quality in the
photographic material in the northern hemisphere.
Valotto et al. (1997) 
suggest a limiting 
apparent magnitude $m=19.4$  for statistical
analysis in COSMOS Survey which minimizes 
errors due to misclassification  of stars, galaxies plate 
variations, etc. 
The rms difference between CCD and
COSMOS magnitudes is $\sigma \approx 0.25$ $mag$ 
(\cite{val97}, \cite{lov92}, 
\cite{rou96}).
This rms value has been used in the Monte-Carlo estimates of
errors described in section 4.

\vskip 15pt
\centerline{TABLE 1 - Group Properties}
\vskip 10pt
\placetable{tbl-2}
\begin{tabular}{@{}ccccl@{}}
\hline
\hline
RA [$\arcdeg$]  & Dec [$\arcdeg$] & $N_{gal}$  & d $[km~ s^{-1}]$ & Reference \\
\hline
   2.7312 & -33.5381  &    &  6488 &  MarkIII    \nl
   9.2596 & -27.3823  &    &  6427 &  MarkIII    \nl
   9.3742 & -20.1418  &  3 &  3474 &  MCL85      \nl
   9.4075 & -46.5883  &    &  3109 &  MarkIII    \nl
  13.8497 & -45.3011  &    &  6173 &  MarkIII    \nl
  14.1764 & -52.9298  &  3 &  7284 &  MCL19      \nl
  17.1555 & -46.2836  &  5 &  6621 &  MCL33      \nl
  17.3560 & -35.8672  &  3 &  6255 &  MCL54      \nl
  18.0305 & -31.7627  &    &  5951 &  MarkIII    \nl
  18.5607 & -33.2295  &    &  6011 &  MarkIII    \nl
  18.6351 & -32.1191  &  7 &  5345 &  MCL61      \nl
  19.8244 & -44.2044  &  4 &  6614 &  MCL36      \nl
  19.9841 & -33.9223  &    &  3187 &  MarkIII    \nl
  20.1977 & -44.8500  &    &  7091 &  MarkIII    \nl
\end{tabular}
\centerline{TABLE 1 - Continued}
\vskip 10pt
\begin{tabular}{@{}ccccl@{}}
\hline
\hline
  21.6424 & -23.2313  &  3 &  7350 &  HC11       \nl
  22.1155 & -35.7583  &  3 &  5308 &  MCL55      \nl
  23.8488 & -39.2949  &  3 &  5309 &  MCL47      \nl
  24.4133 & -27.4186  &    &  5713 &  MarkIII    \nl 
  24.8495 & -42.7212  &    &  6012 &  MarkIII    \nl
  26.9436 & -35.0977  &    &  4715 &  MarkIII    \nl
  29.2886 & -50.4484  &    &  5546 &  MarkIII    \nl
  31.3328 & -55.1617  &  3 &  5725 &  MCL17      \nl
  32.4185 & -41.6438  &  3 &  5254 &  G50        \nl
  32.3422 & -22.5234  &    &  3538 &  MarkIII    \nl
  32.8671 & -23.2635  &    &  4668 &  MarkIII    \nl
  33.0409 & -39.4327  &  3 &  4924 &  MCL46      \nl
  33.2765 & -32.0978  &    &  3829 &  MarkIII    \nl
  37.3372 & -31.4779  &  3 &  4393 &  MCL64      \nl
  37.4580 & -42.9617  &  3 &  5027 &  MCL41      \nl
  40.8292 & -55.2613  &  3 &  5398 &  G74        \nl
  41.3068 & -17.6733  &  3 &  7232 &  MCL87      \nl
  41.6544 & -24.9411  &  3 &  6416 &  MCL75      \nl
  41.8712 & -55.3755  &  3 &  5483 &  MCL16      \nl
  42.5540 & -31.2784  &    &  5137 &  MarkIII    \nl
  43.2061 & -32.1921  &    &  5597 &  MarkIII    \nl
  44.1086 & -18.5158  &    &  3577 &  MarkIII    \nl
  50.4756 & -13.6459  &  7 &  9510 &  HC26       \nl
  50.5104 & -24.7825  &    &  5243 &  MarkIII    \nl
  52.0631 & -21.2036  &    &  2179 &  MarkIII    \nl
  59.2783 & -25.3789  &    &  3782 &  MarkIII    \nl
  64.1942 & -44.2942  &    &  4552 &  MarkIII    \nl
 321.1159 & -40.4012  &  4 &  4983 &  MCL45      \nl
 322.5890 & -39.5620  & 18 &  5145 &  G445       \nl
 324.9505 & -39.9815  &    &  4785 &  MarkIII    \nl
 328.1029 & -28.8809  &    &  6449 &  MarkIII    \nl
 329.1513 & -34.7459  &    &  2625 &  FGCP244    \nl
 330.6722 & -34.3896  &  3 &  4479 &  G451       \nl
 330.8851 & -32.1050  &    &  2842 &  MarkIII    \nl
 331.3571 & -18.9131  &    &  2749 &  MarkIII    \nl
 331.5865 & -27.8394  &  3 &  6770 &  MCL69      \nl
 332.3014 & -27.7760  &  4 &  7140 &  HC91       \nl
 332.3542 & -32.8460  &    &  3760 &  MarkIII    \nl
 333.2208 & -30.0362  &  3 &  4352 &  G454       \nl
 333.7056 & -27.4679  &  3 &  5190 &  MCL70      \nl
 335.2614 & -21.8241  &    &  5912 &  MarkIII    \nl
 335.4913 & -25.5694  &  7 &  4805 &  G457       \nl
 336.5680 & -28.2023  &    &  6551 &  MarkIII    \nl
 336.5773 & -14.9864  &  5 &  5066 &  G458       \nl
 337.9965 & -27.0941  &    &  3435 &  MarkIII    \nl
 338.1969 & -19.9765  &    &  7474 &  MarkIII    \nl
 341.4786 & -22.9033  &  3 &  3039 &  MCL76      \nl
 341.5522 & -39.8120  &    &  2555 &  MarkIII    \nl
 341.7496 & -22.6184  &    &  3369 &  FGCP251    \nl
 342.9665 & -20.1329  &    &  3184 &  MarkIII    \nl
 351.5385 & -35.3763  &    &  3177 &  MarkIII    \nl
 352.9783 & -46.7819 &    &  3267 &  MarkIII    \nl
 354.5960 & -47.6564  &  3 &  2781 &  MCL30      \nl
\hline
\hline

\end{tabular}

\vskip 5pt

Table 1 lists the corresponding angular
positions (J2000.0), published number of members 
$N_{gal}$ when available and distances
d (redshift independent estimates for MarkIII groups, and d=cz otherwise) in km/sec.\\
FGCP: Fouque et al., \ 1992;
G: Garcia 1993;
HC: Hickson 1982;
MCL: Maia et al., \ 1989;
MarkIII: Willick et al., \ 1997.

\section{ANALYSIS AND RESULTS}

\subsection{Galaxy Counts and Background Subtraction.}

We compute the number of galaxies brighter than a limiting absolute
magnitude $M_{\lim }$ within a projected radial distance $r$ from the
centers of the groups. 
We have applied a $K-$correction term of the form $K=2z$ (\cite{esf88}). 
The group projected radius $r$ was fixed at $
0.5$ $h^{-1}$ $Mpc$ where Hubble constant is 
$H_0=100$ $h$ km $s^{-1}$ Mpc$^{-1}$ similar to the adopted
group radius in other studies (see for instance \cite{gel83}, \cite{gar93}).

We define a mean local background around each group 
in order to decontaminate the galaxy counts. 
This mean local background is defined as the number density of galaxies in the 
same range of
apparent magnitudes in a ring at projected radii $R_{1}<r<R_{2}$. 
According to \cite{val97} the stability of the results does not depend
crucially on the adopted radius for 
background correction provided that 
the decontamination ring is well beyond the
average projected radius of groups and small enough in order to take into
account local variations of the projected galaxy density due 
to patchy
galactic obscuration, large scale gradients in the galaxy catalog, etc. The
counts of galaxies for each group are binned in magnitude intervals of $
0.5$ $mag$ which is larger than the photometric errors. 
We subtract the corresponding mean background correction to
each magnitude bin to compute the contribution from each group to the
LF. The total decontaminated number of galaxies in the range
$-22 < B_{j} < -16$ is $N_{tot} = 1280 $ corresponding to an average
number of galaxies $ n \simeq 20$ per group which may be compared
to $ n \simeq 100$ in clusters (see \cite{val97}, Table 2). 
We notice that in spite of this large difference in the number of galaxies 
,groups and clusters have comparable mean galaxy densities due to
 the radii adopted (0.5 and 1.5 $h^{-1}$ Mpc respectively).

\subsection{Results and Error Estimates}

Figure 2 shows the galaxy group
LF of the total sample of groups with $M_{lim}=-16$. 
Error bars in this figure correspond to the Monte-Carlo
determination of errors discussed below which provide a  
reliable estimates of the uncertainties.

Errors in the  determination of the galaxy LF without
individual galaxy distances arises from uncertainties in the
decontamination of foreground and background galaxies. 
\cite{val97} found a good agreement between the galaxy cluster LF determinations
from different background corrections although
fluctuations in the background counts 
contribute significantly to errors. 
We have also considered the propagation of photometric errors   
in COSMOS
magnitudes in order to provide suitable error estimates of the galaxy LF in
groups. We use a   
 Monte-Carlo algorithm that takes into account  
the observed scatter
between 
CCD and COSMOS magnitudes 
through random Gaussian errors with dispersion
 $\sigma=0.26$ mag. added to galaxy magnitudes 
as well as errors associated to the different
assignment of background corrections. 
For each group we obtain different background corrections by counting galaxies
in circles of projected radius $r=0.5$ h$^{-1}$ Mpc centered  at
distances 
2 h$^{-1}$ Mpc$<r_1<$4 h$^{-1}$ Mpc from the groups.
We calculate the resulting LF in 50 random realizations
which include magnitude errors and background fluctuations. 
Our Monte-Carlo estimates of errors in figure 2 
correspond to the rms values of the
relative number frequency of galaxies in each absolute magnitude interval
$M-M+\Delta M$. 

\placefigure{fig-2}
\input epsf
\def\figureps[#1,#2]#3.{\bgroup\vbox{\epsfxsize=#2
    \hbox to \hsize{\hfil\epsfbox{#1}\hfil}}\vskip12pt
    \small\noindent Figure#3. \def\par{\endgraf\egroup\vskip12pt}}
\figureps[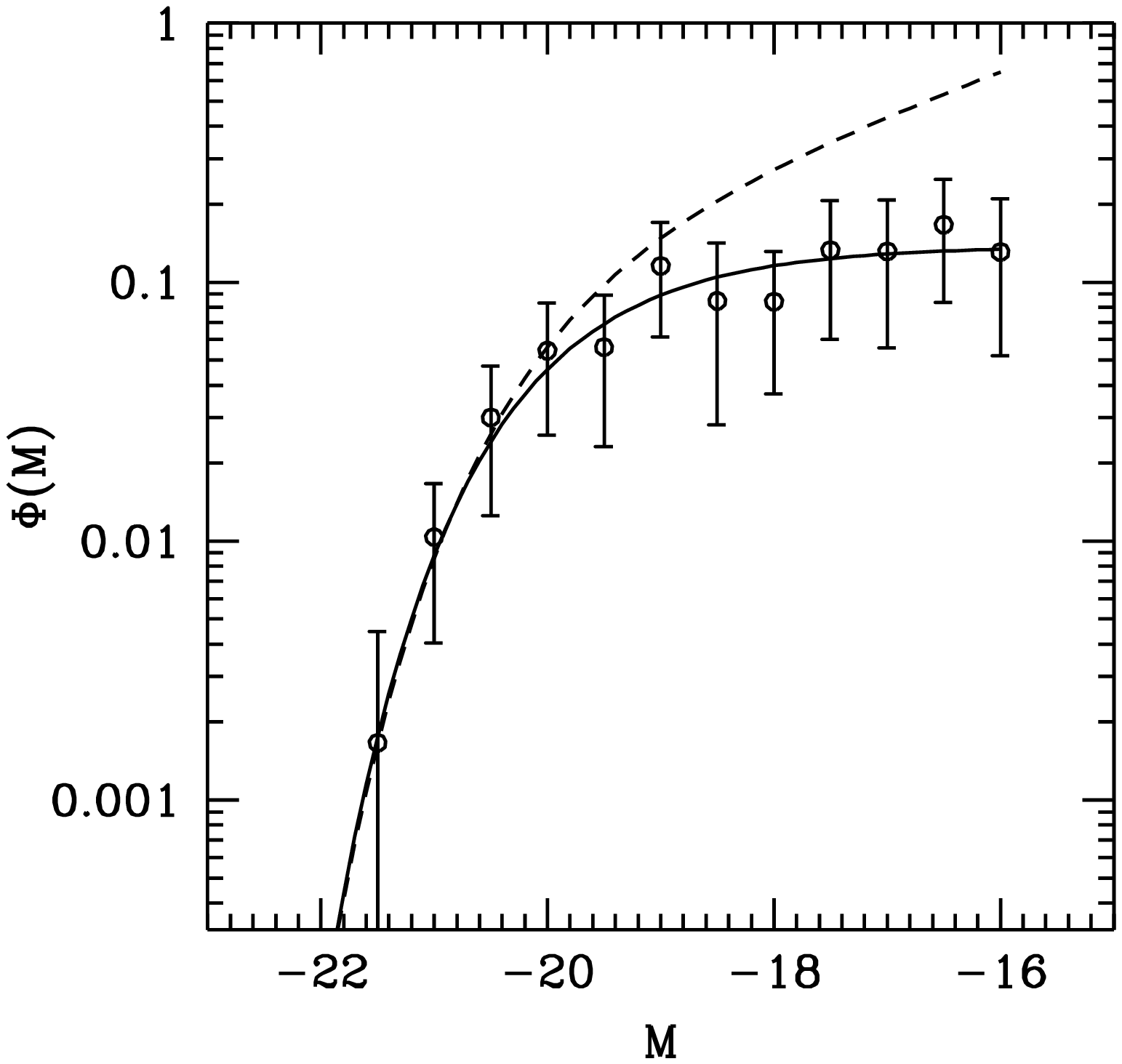,1.00\hsize] 2. Uncorrected luminosity function of galaxies in groups. 
Solid line corresponds to the Schechter best fit with parameters
$M^{*}= -19.9$ and $\alpha = -1.0$. Dashed line correspond to the


In order to provide a suitable LF fit, we have adopted a Schechter function
model $\phi (L)dL=const\times (L/L^{*})^\alpha e^{-L/L^{*}}d(L/L^{*})$
(\cite{sch76}). We have applied a maximum-likelihood estimator using a $
\chi ^2-$minimization procedure developed by Levemberg and Marquard, (see
\cite{pre87}) which minimizes the difference
$$\chi ^2\equiv \sum_{i=1}^N\left( \frac{\phi _i-\phi (L_i;const,\alpha
,L^{*}) }{\sigma _i}\right) ^2 $$
where $\phi _i$ is the relative frequency of galaxies corresponding to the $
i^{th}$ luminosity bin and $\sigma _i$ is its associated uncertainty. The
problem is reduced to the derivation of the three parameters $const,L^{*}$ and $
\alpha $ which minimizes $\chi ^2$. This method deals with errors 
giving a reliable set of fitting parameters provided the errors are
representative of the true uncertainties involved. In our calculations we
have considered errors derived through the Monte-Carlo estimates
described above which we argue, give confident estimates of errors. 
In figure 2 is shown (solid line) the Schechter fit
derived by the maximum-likelihood estimator with parameters $\alpha =-1.0$, $
M^{*}=-19.9$ where can be seen a very good agreement 
between observations and the Schechter fit. 

We have considered the effects on the Schechter function
fitting parameters due to magnitude binning and 
photometric errors through a Monte-Carlo technique. 
We simulate a Schechter function LF  
with parameters corresponding to the group galaxy LF ($\alpha =-1.0$, 
$ M^{*}=-19.9$) 
and the cluster galaxy LF ($\alpha=-1.4$, $ M^{*}= -20.0$) provided by \cite{val97}.
For both cases we have considered Gaussian photometric errors with dispersion
$\sigma 0.25$ mag) and we have taken into account the adopted binning 
(0.5 mag, and 0.3 mag in groups and clusters respectively).
In both cases we find a resulting $M^{*}$ parameter $\simeq 0.3$ mag lower
than the input values while the parameter $\alpha$ remains unchanged.
The corresponding corrected fitting parameters for the group and
cluster galaxy LF are $\alpha =-1.0$, 
$ M^{*}=-19.6$) and $\alpha=-1.4$, $ M^{*}= -19.7$ respectively.

A summary of the results obtained is given in Table 2.
For comparison are also listed in this table the resulting
Schechter fitting parameters of the corrected field galaxy LF
(\cite{lov92}).
It can be seen in this table the importance of environment on the shape
of the LF. The group galaxy LF has a flat faint end  
consistent with the field galaxy LF in contrast to clusters 
where an important rise in the number of low luminosity galaxies is present. 

\vskip 15pt
\centerline{TABLE 2 - Summary of Statistical Results}
\vskip 10pt
\placetable{tbl-2}
\begin{tabular}{@{}lcc@{}}
\hline
\hline
Samples & $\alpha$ & $M^{*}$ \\
\hline
\hline
Groups & -1.0 $\pm$ 0.2  & -19.6 $\pm$ 0.2\nl
Rich Clusters & -1.5 $\pm$ 0.1  & -19.7 $\pm$ 0.1\nl
Poor Clusters & -1.2 $\pm$ 0.1 &  -19.6 $\pm$ 0.1\nl
Field Galaxies & -1.0 $\pm$ 0.15 &  -19.5 $\pm$ 0.1\nl   
\hline
\hline
\end{tabular}
\vskip 15pt

We show in figure 3 the corresponding $
1\sigma $ and 2$\sigma $ level error contours of corrected galaxy  LF in
groups 
 (contours $\chi$ $^2-\chi _{ML}^2$) in the $\alpha $ $-M^{*}$ plane .
In this figure the corrected cluster and field galaxy LF fitting parameters
given by \cite{val97} and \cite{lov92}
are also displayed.

\placefigure{fig-3}
\input epsf
\def\figureps[#1,#2]#3.{\bgroup\vbox{\epsfxsize=#2
    \hbox to \hsize{\hfil\epsfbox{#1}\hfil}}\vskip12pt
    \small\noindent Figure#3. \def\par{\endgraf\egroup\vskip12pt}}
\figureps[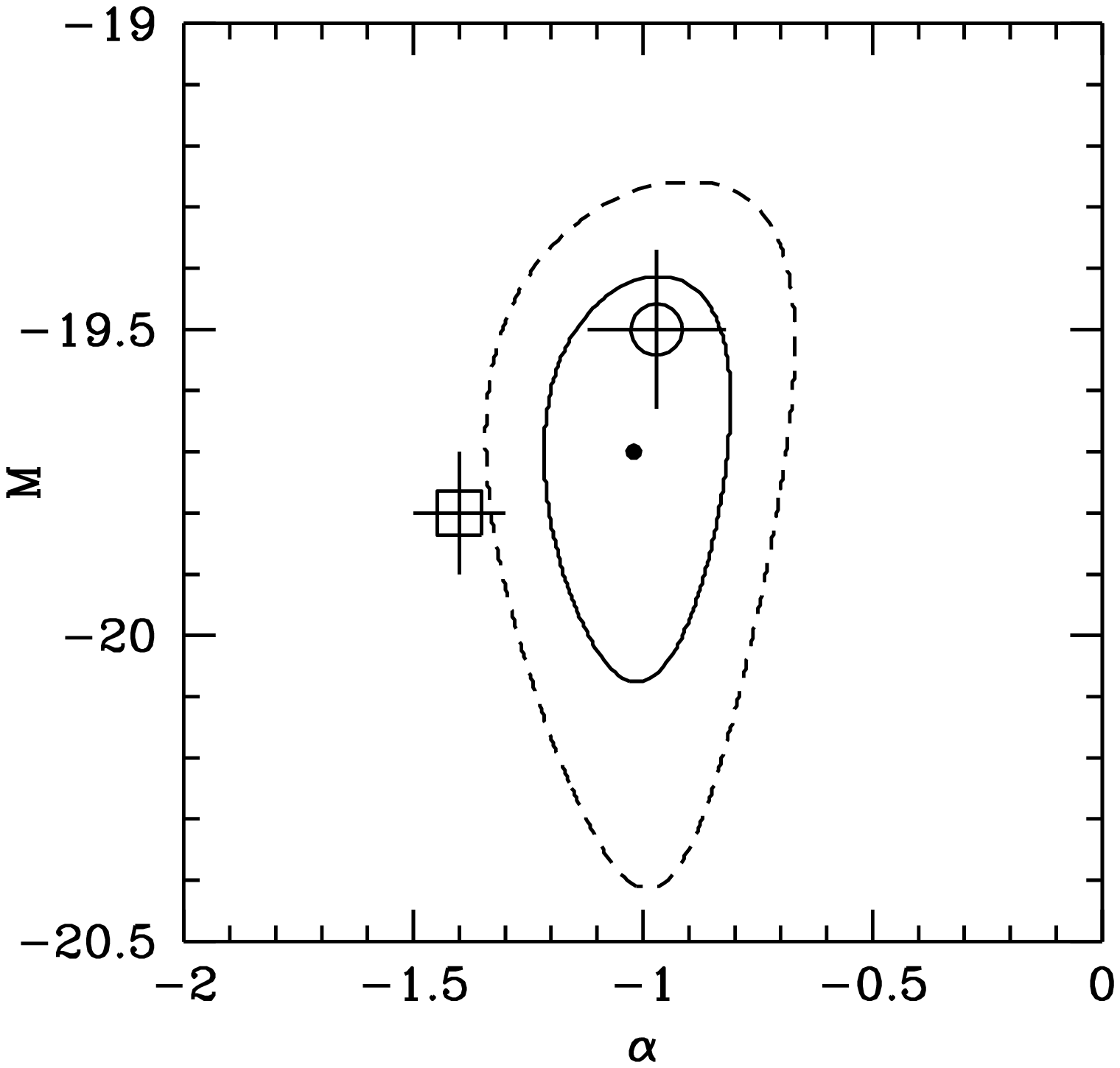,1.00\hsize] 3. Error contours 1$\sigma$ (solid line)
and 2$\sigma$ (dashed line) of the Schechter function parameters
$\alpha$ and $M^*$ for the corrected galaxy LF in groups. 
The open circle and square correspond to the corrected cluster and field
galaxy LF (Valotto et al. 1997 and Loveday et al. 1995).

\section{CONCLUSIONS}

We have applied statistical analyses to calculate the mean galaxy 
luminosity function in groups. We have used COSMOS survey galaxies 
and group catalogs with either available
distances or redshifts.    
Schechter functions provide good fits 
to the galaxy LF of the sample of groups analyzed.
The best fitting parameters for our total sample of 66 groups 
are $\alpha =-1.0\pm
0.2$, $M^{*}=-19.6\pm 0.2$. This value of the
 $\alpha $ parameter contrasts with that corresponding to a single Schechter fit in clusters 
$\alpha \approx -1.4$ (see for instance \cite{fer91}, \cite{val97})
indicating an
important relative excess of faint galaxies in clusters.
The dependence of the galaxy LF on cluster
richness was explored by \cite{val97} 
finding that
poor clusters show a flatter galaxy LF than more massive
systems ($\alpha \simeq -1.2$ in contrast to $\alpha \simeq -1.5$).
\cite{lc97} results
show a flat galaxy LF ($\alpha \simeq -1.0$)
in a subsample of dynamically relaxed
systems  which may  be a strong indication that evolutionary process may disrupt dwarf galaxies
contributing to the formation of cD haloes in these systems. 
However, \cite{driver} in their analyses of the galaxy LF of the rich 
(R=3) cD-dominated cluster A963 a large excess of faint galaxies is found
which indicates the non universality of this phenomenon. Similar large
excess of faint galaxies are reported by \cite{tren97} for 4 rich clusters
(R=2-4) indicating that the dependence
of the faint excess on different physical parameters of clusters
deserve further analysis.  
The importance of the cluster merging history 
and the gas-dynamical evolution within
these systems should be addressed before firm conclusions can be derived from
the analysis of the observations.

Our comparison with \cite{val97} results 
have the advantage that both have used similar 
methods of analysis, same galaxy catalog and background 
subtraction procedure and therefore it provides a reliable confrontation of the
mean galaxy LF in clusters and groups.
In spite of the fact that our sample galaxies 
correspond to moderate galaxy overdensities
we find no evidence for an 
excess of faint galaxies relative to the field as it is found for the 
mean galaxy LF in clusters.
Although galaxy mutual interactions are expected to play an 
important role in groups 
our results indicate that at these moderate galaxy density enhancements
the galaxy luminosity function is not significantly different from the field.

\acknowledgments

We acknowledge Dr. MacGillivray for kindly permitting the use of the COSMOS
Survey. 
We thank the Referee for useful comments and suggestions which
greatly improved the original version of this paper.
This work was partially supported by the Consejo de
Investigaciones Cient\'{\i}ficas y T\'ecnicas de la Rep\'ublica Argentina,
CONICET, the Consejo de Investigaciones Cient\'{\i}ficas y Tecnol\'ogicas de
la Provincia de C\'ordoba, CONICOR, and Fundaci\'on Antorchas,
Argentina.





\end{document}